\documentstyle{article}
\begin{document}
\rightline{Nuovo Cimento D 19 (Oct. 1997) 1627}
\rightline{physics/9610017}
\begin{center}
{\bf Note on the dynamic instability of microtubules}\\

H.C. Rosu
\footnote{e-mail: rosu@ifug3.ugto.mx}\\[2mm]

Instituto de F\'{\i}sica - IFUG,
Apdo Postal E-143, 37150 Le\'on, Gto, M\'exico\\
\end{center}


\begin{center}
{\bf Abstract}

If the dynamic instability of microtubules follows a gamma
distribution then one can associate to it a Cantor set.

\end{center}


Microtubules (MTs), the main protein polymeric filaments of the cell
cytoskeleton, are a well-defined biological system where methods in
condensed matter, statistical mechanics and the theory of complex systems
have been applied. MT dynamics plays an important role in many fundamental
cellular processes, such as cell division and cell motility.
A peculiar intrinsic dynamics of MTs consisting in extended
growth (rescues) and shrinkage (catastrophes) phases of variable duration
with rapid switching
between these two phases has been known since about a decade \cite{mk} and is
the subject of many investigations. This spontaneous assembly-disassembly
process has been called {\em dynamic instability} (DI) \cite{mk}. It is not
clear why MTs should have such a behavior and what is its true origin.
There are currently several kinetic models trying to explain DI,
both from the side of biologists \cite{biol} (based on specific features of
the MT hydrolysis involving the GTP and GDP units)
and from the side of physicists \cite{phys} (based on purely one-step
stochastic processes of constant rate).

Recently, Odde, Cassimeris, and Buettner \cite{ocb}
have found a high probability of fitting the
published MT length life histories to a gamma distribution
($f(t)=\theta ^{r}t^{r-1}e^{-\theta t}/\Gamma (r)$, where $r$ and $\theta$
are the shape parameter and the frequency parameter, respectively,
of the distribution) by the
Kolmogorov-Smirnov test, for both growth and shrinkage. Their result is
not firmly established because the conclusion is based on a small
number of phase times (14 growth times and 12 shrinkage times). The
exponential distribution still cannot be ruled out. As they remark,
other nonnegative probability distributions (Weibull, lognormal, or beta ones)
may also be appropriate. However, the property of memory makes all these
distributions essentially different from the exponential one.
Citing the textbook of Olkin, Gleser, and Derman \cite{ogd},
Odde and collaborators argue that a gamma distribution implies a series of
first-order steps from the growth phase to the shrinkage one, with the
number of steps given by the shape parameter $r$ of the gamma distribution.
For the plus end data, they have found $r=3$, and thus a series of three
first-order transitions each of constant rate $\theta$
($\approx$ 1.7 min$^{-1}$). This would mean two intermediate metastable
states. In their words, ``each of these transitions could potentially
represent key chemical or physical events occurring in the MT".

The purpose of this note is to point out another connection of the gamma
distribution (actually going in the same direction as the aforementioned
surmise of Odde {\em et al}) that makes such a distribution
very appealing in the case of MTs, if proven. The connection may be found
in the review paper of Carruthers and Shih on the phenomenology of
hadronic multiplicity distributions \cite{cs}. The point is that
the asymptotic form of negative binomial (already written down by Mandel in
1959) is a special case of the
gamma distribution (i.e., the case $r=\theta$). Moreover,
Carruthers and Shih enumerate possible origins of the negative binomial
distribution. The last origin in their list of six items is as a realization
of self-similar Cantor set structures. As they commented, the negative
binomial distribution can be put into one-to-one correspondence with
self-similar sequences whose limits lead to the (triadic) Cantor sets.
The message of all these connections might well be of great relevance for
the MT dynamics, because Cantor sets are directly related to multifractal
scaling properties \cite{cmf} and these would mean self-similar features
in the MT dynamic instability. In case the gamma fits of Odde {\em et al}
are confirmed, further studies of the nature of the energetic
self-similar trees will be necessary.

\begin{center}      ***  \end{center}

I acknowledge a stimulating discussion in Houston and correspondence
with Dr. David Odde. He was also very kind in sending me his papers.
I thank Prof. Carruthers for promptly sending me his review with
Shih \cite{cs}.


\end{document}